\documentclass[twocolumn]{jpsj3} 
\usepackage[usenames]{color}
\usepackage{txfonts}
\newcommand{\crb}{CeRh$_{3}$B$_{2}$}

\def\runauthor{Kunihiko \textsc{Yamauchi}, 
Akira \textsc{Yanase} 
and Hisatomo \textsc{Harima}}
\title{%
Bandstructure and Fermi Surfaces of CeRh$_{3}$B$_{2}$
}

\author{%
Kunihiko \textsc{Yamauchi}$^1$\thanks{Present address: CNR-INFM, CASTI, Regional Lab, I-67010 Coppito (LÕAquila), Italy},
Akira \textsc{Yanase}$^1$ and 
Hisatomo \textsc{Harima}$^2$}

\inst{%
$^1$ISIR-SANKEN, Osaka University,  8-1, Mihogaoka, Ibaraki, Osaka 567-0047, Japan \\
$^2$Department of Physics,  Kobe University, 1-1, Rokkodai, Nada, Kobe 657-8501, Japan  \\
}

\recdate{\today}

\abst{%
The electronic bandstructure and the Fermi surfaces of ferromagnetic CeRh$_{3}$B$_{2}$
 are calculated by using FLAPW and LSDA+{\it U} method. 
As assuming  several kinds of the ground state to describe the $4f$ electronic state,  
 we propose a fully orbital- and spin-polarized state $\left| {l_{z}=0, s_{x}=1/2} \right\rangle$ as the ground state, instead of the conventional $LS$-coupled CEF ground state,  generally expected in typical $4f$ compounds.
This is supported by the fact that both the observed magnetic moment and the observed dHvA frequencies are well explained by the calculated electronic structure and the Fermi surfaces.
The unconventional ground state is stabilized by the strong $4f$-$4f$ direct mixing
between the neighbored Ce atoms along the extremely small distance along the $c$-axis in the hexagonal crystal cell.  
}

\kword{%
$\rm CeRh_3B_2$, LSDA+{\it U}, Fermi surface
}

\begin{document}
\maketitle

\section{Introduction}

The trivalent cerium compounds show the versatile features such as 
the heavy fermion behavior,\cite{heavy1, heavy2} 
the anisotropic superconductivity,\cite{heavy1} the complicated magnetic phase,\cite{magphase1, magphase2} and the multipole order.\cite{multipole} 
These phenomena demonstrate both the localized and itinerant behavior of $4f$ electrons.
Among the ferromagnetic cerium compounds with other nonmagnetic elements, 
ternary cerium boride CeRh$_{3}$B$_{2}$  
has the highest Curie temperature $T_{\rm C}=$ 120 K \cite{cerh.curie1}, 
while the usual magnetically ordering temperature is only 1-10 K in other cerium compounds. 
In contrast to the highest $T_{\rm C}$,  it is notable that the saturated magnetic moment in CeRh$_{3}$B$_{2}$ 
is remarkably small.
The magnetic measurement done by Galatanu $et. al.$ \cite{cerh.galatanu} at 2K has shown 
that the saturated magnetization shows the strong anisotropy  in the hexagonal crystal structure: 
 0.451$\mu_{\it{B}}$ along [10\=10] ;
 0.447$\mu_{\it{B}}$ along [11\=20] ; 
 0.04  $\mu_{\it{B}}$ along [0001] direction
 so that they have concluded that the easy axis is the [10\=10] direction. 
This small values of the magnetic moment ($\approx$0.45$\mu_{\it{B}}$)  are considerably smaller than the value ($\approx$ 1.0 $\mu_{\it{B}}$/Ce)\cite{cerh.kasuya.okabe} of general ferromagnetic Ce compounds.
In order to explain the unusual ferromagnetic behavior, 
a Ce 4$f$ itinerant ferromagnetic model \cite{cerh.f.it}, 
a Rh 4$d$ itinerant ferromagnetic model \cite{cerh.d.it} and 
Ce 4$f$ localized ferromagnetic models \cite{cerh.f.loc1,cerh.f.loc2,cerh.f.loc3,cerh.f.loc4,cerh.f.loc5,cerh.f.loc6,cerh.galatanu} 
have been proposed so far.

In the 4$f$ localized model, 
the ground state of the Ce $4f$ electron is described by the crystalline electric field (CEF) splitting and the relatively stronger spin-orbit splitting.  
In the hexagonal symmetry, 
the six-fold degenerate $\left| {j =  5/2} \right\rangle$ level splits into three doublets, where
the ground states is described by $\left| {j_z  =  \pm 1/2} \right\rangle$. 
However, this scheme fails in the explanation of the observed small magnetic moment 
because the doublet ground state $\left| {j_z  =  \pm 1/2} \right\rangle$ is expected to cause very large saturated moment 
of 1.6 $\mu_{\it B}$ in the basal plane.
Therefore, the 4$f$ localized model with considering a CEF ground state hardly explain the observed small magnetic moment. 

On the other hand, the de Haas-van Alphen (dHvA) measurement, observed for LaRh$_{3}$B$_{2}$ and ferromagnetic CeRh$_{3}$B$_{2}$ \cite{cerh.dhva},  strongly supports the $4f$ localized model.
This observation has suggested that 
the contribution of the $4f$ electrons to the Fermi surfaces  is small ($i.e.$ the $4f$ electron  is well localized)
because the topology of the Fermi surfaces in CeRh$_{3}$B$_{2}$ is similar to that of LaRh$_{3}$B$_{2}$, 
which is consistent with the theoretical result \cite{larh.fermi} calculated by a full potential LAPW (FLAPW) method. 
Therefore,  the localized $4f$ model can be applied to explain the Fermi surfaces, 
however, as mentioned above, 
there is inconsistency in the localized model in the magnetic moment.

In this paper, we report the theoretical study of the electronic structure of \crb, aiming at clarifying the property of the localized $4f$ electron.
First we will explain the detail of the  method of the calculations, then 
 we will discuss the ground state of the $4f$ electron, which leads to the particular magnetism and the Fermi surfaces.

\section{Computational Details}

We have performed a bandstructure calculation within the local density approximation (LDA)
in the density-functional framework  
using the full potential linearized augmented plane-wave (FLAPW) formalism\cite{flapw}.
We used TSPACE and KANSAI-99 program codes for this calculation.

The scalar relativistic effects are taken into account for all electrons and
the spin-orbit interactions are included for the valence electrons inside the Muffin-tin spheres 
as in a second variational procedure.\cite{koellingharmon}  
In the perturbation Hamiltonian for the spin-orbit coupling (SOC),  the radial potential of the spin-offdiagonal element is 
substituted as an average of the potential for the spin-up and down state.\cite{kubler}
The SOC Hamiltonian term is unitary transformed by using the spin-rotation matrix, which defines the quantization axis  so that the calculated electronic state is dependent on the direction of the magnetic moment. The net orbital moment is induced by the SOC which breaks the time-reversal symmetry. 

In order to treat carefully the strong correlated  4$f$ state, 
the effective L(S)DA+$U$ potential \cite{anisimov} is applied to improve the LSDA calculation.
For the $4f$ electron system, where the spin-orbit coupling is involved,
the density matrix is described as a spin- and orbital-dependent ($i.e.$ $14\times14$) matrix, considering  the spin-flip element\cite{harima.ldau}. 
The effective potential is introduced into the calculation as a second variation, together with the SOC Hamiltonian.
In this procedure, although the spin- and orbital-dependent density matrix converges self-consistently with the wavefunctions,
the resulted $4f$ state is strongly dependent on the initial density matrix with a certain value of $U$.
In other words, the ground state of the $4f$-state is determined  {\it a posteriori}. 
Based on the same method, a metamagnetic transition of Fermi surfaces of CeRu$_{2}$Si$_{2}$ has been successfully investigated.\cite{suzuki} 
In the following sections, we will first report the bare LSDA result and later the LSDA+$U$ results with considering the several $4f$ ground states. 

\begin{figure}[htbp]
 \begin{center}
 \includegraphics[width=7cm,clip]{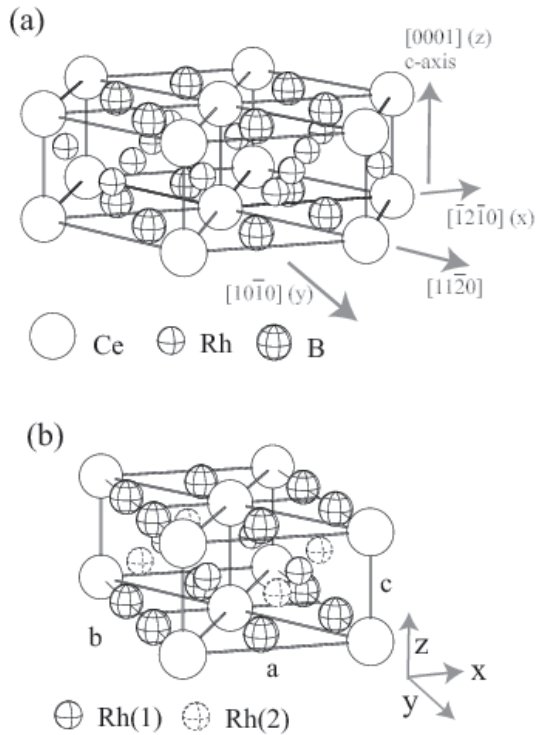}
 \end{center}
 \vspace{-4mm}
 \caption{Crystal structure of CeRh$_{3}$B$_{2}$;  
	(a) The hexagonal structure containing three unit cells and (b) the convention unit cell of the base-centered orthorhombic structure with the lower symmetry in the ferromagnetic state. The Cartesian coordinates used in the calculation are also shown.}
 \label{fig:cerh.crys}
\end{figure}

{\crb} crystallizes in a hexagonal CeCo$_{3}$B$_{2}$ crystal structure ($D_{6h}^{1}$ space group), shown in Fig.  \ref{fig:cerh.crys}(a).
It should be emphasized that  the hexagonal unit cell is used only for the calculation of the non-magnetic state
whereas the less symmetrical orthorhombic unit cell is used for the ferromagnetic state.
This is because  the  symmetry of the electronic state is broken by the SOC, which couples the freedom of spin and orbital, so that the reduced symmetry is dependent on the  direction of the collinear magnetic moment.
As taking into account of the previously reported magnetic measurement \cite{cerh.galatanu}, which shows that  the magnetic moment is oriented in the basal plane with the small in-plane magnetic anisotropy,
we have adopted the base-centered orthorhombic unit cell ($D_{2h}^{19}$ space group, shown in Fig. \ref{fig:cerh.crys}(b)),
where the magnetic moment is assumed to be aligned along $x$ ([\=12\=10] or equivalently [11\=20] in the hexagonal structure) axis, and also another setting where the magnetic moment is aligned along the $y$ ([10\=10]) axis is considered, for the comparison. 

The experimentally measured structural parameters \cite{cerh.crysl} are used in the calculation :
$a=5.474$ \AA\ and $c = 3.085$ \AA\ (Note that $b=a$ in the hexagonal cell and $b=\sqrt{2}a$ in the orthorhombic cell).
Muffin-tin radii are set as 
$0.2761a$, $0.2192a$ and $0.1761a$
for Ce, Rh and B sites respectively.  
The core electrons (Xe-core except 5{\it s}$^2$ and 5{\it p}$^6$ for Ce, Kr-core except 4{\it p}$^6$ for Rh, He-core for B) are calculated inside the MT spheres in each self-consistent step.
The LAPW basis functions are truncated at $|{\bf k+G}_{i}|= 5.81 \times 2 \pi /a$, 
corresponding to 409 LAPW functions at the $\Gamma$ point. 
The sampling 150 $k$-points (divided by 6, 6 and 10) are uniformly distributed 
in the irreducible 1/8th of the orthorhombic Brillouin zone.

\section{LSDA result}
\begin{figure}[htbp] 
\begin{center}
\includegraphics[width=7cm]{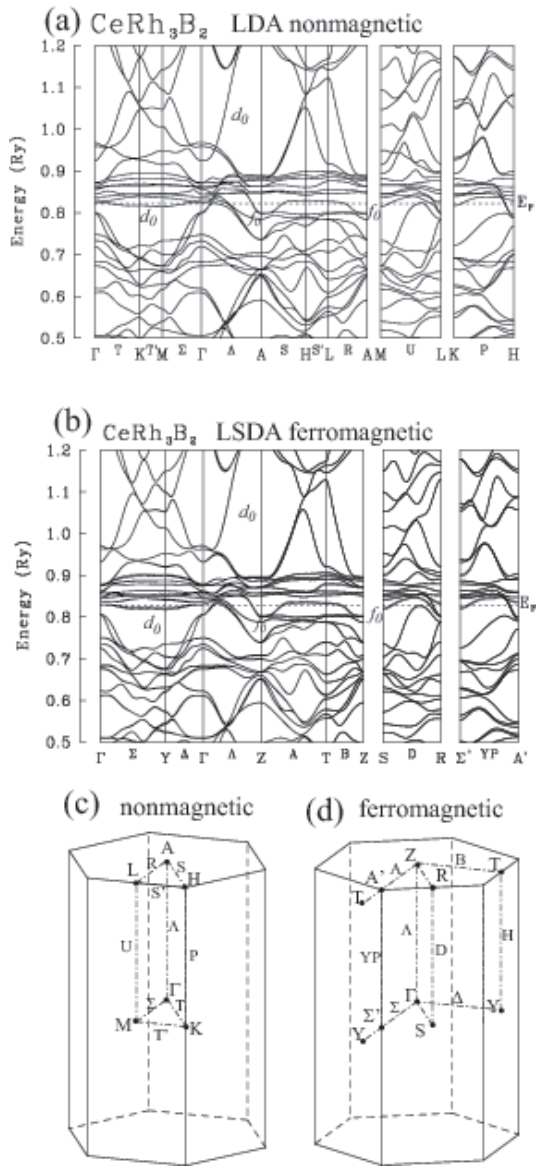}
\end{center}
\vspace{-4mm}
\caption{Bandstructure of CeRh$_{3}$B$_{2}$ for (a) the  nonmagnetic state and (b) the ferromagnetic state calculated in the bare LSDA scheme. The first Brillouin zone for (c) the  nonmagnetic hexagonal lattice and (d) the ferromagnetic orthorhombic lattice is also shown. In (b) and (d),  $\Sigma$' (2/3, 0, 0) and A' (2/3, 0, 1/2) points and YP (2/3, 0, $k_{z}$) line are newly defined.}
\label{fig:cerh.band.lda}
\end{figure}
\begin{figure}[htbp] 
\begin{center}
\includegraphics[width=7cm]{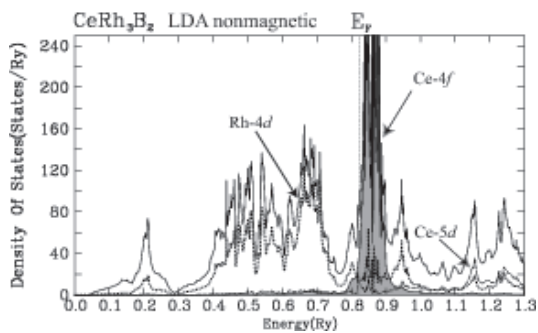}
\end{center}  
\vspace{-4mm}
\caption{
An LDA result of the density of states for nonmagnetic CeRh$_{3}$B$_{2}$. 
Partial density of states of Ce-$f$ (highlighted), Ce-$d$ (dashed line) are Rh-$d$ (dotted line) states are indicated. }
\label{fig:cerh.dos.nonmag.lda}
\end{figure}

Before we proceed to the LSDA+$U$ calculation, let us focus on  the bare LSDA result,
where the $4f$ state is treated as an itinerant state. 
Considering the ferromagnetic state, the spin magnetization is set to be parallel to the $x$ axis 
([11\=20] direction in the hexagonal structure as in Fig 1(a)).
In this study, we use the definition  that the spin and orbital quantization is described with respect to $z$ axis ([0001] axis in the hexagonal structure). 
Therefore, the magnetic quantum number $m$ of the spherical harmonics of the LAPW basis corresponds to the $z$-projected orbital moment $\langle l_{z}\rangle$.

Figure \ref{fig:cerh.band.lda} shows the calculated bandstructure for CeRh$_{3}$B$_{2}$ nonmagnetic and ferromagnetic states within the L(S)DA scheme. 
They show the similar aspects  except for the tiny spin-splitting of the $f$ bands near the Fermi energy
in the ferromagnetic bandstructure, where the up- and down- spin states are mixed by SOC. 
In the energy region which is shown in the figure, the bandstructure mainly consists of Rh-$d$ states below the Fermi energy and the Ce-$d$ states above the Fermi energy. 
Figure \ref{fig:cerh.dos.nonmag.lda} shows the density of states for the nonmagnetic state. 
The main components of density of states at the Fermi energy are 
Ce-$f$ (48\%), Rh-$d$ (18\%) and Ce-$d$ (17\%) states. 
Due to  the small polarization of both the spin and the orbital states,
the value of the net magnetic moment along $x$ axis is  negligibly small as 0.07 $\mu_{B}$ 
in the ferromagnetic state. 
Therefore, the bare LSDA calculation, based on the itinerant $f$ model, fails to reproduce the observed magnetic moment. 
In addition, the calculated Fermi surfaces (not shown) 
are completely different from what expected from the observed dHvA observation\cite{cerh.dhva}, 
which is similar to those of prototype  LaRh$_{3}$B$_{2}$. 
This is because the calculated $f$ bands have itinerant nature, as crossing the Fermi level.

In order to improve the LSDA result, we introduce the effective LSDA+$U$ potential into the Ce-$4f$ state.
In this procedure, the effective potential shifts down 
the occupied $f^{1}$ level by $-U/2$ and
shifts up the unoccupied $f^{13}$ levels by $+U/2$ with respect to the original LSDA levels.
The occupied $f^{1}$ level in the LSDA bandstructure is labeled as ``$f_{0}$'' in Fig. \ref{fig:cerh.band.lda}. 
This  state has been already discussed by Takegahara {\it{et al.}} within 
the APW calculations \cite{cerh.apw}.
They have pointed out that due to the strong $dd\sigma$ and $ff\sigma$ mixing along the extremely short Ce chain along the $z$ axis, 
the bottom of the Ce-$4f_{0}$ ($\equiv Y_{30}$) and  $5d_{0}$  ($\equiv Y_{20}$) bands are shifted downward largely with respect to other $f$ and $d$ bands and show the large dispersion in energy. 
This mechanism is also explained by  the CEF point charge model: 
the $f_{0}$  wave function tends to extend to the nearest neighbor positively charged Ce$^{3+}$ ions.
In our LSDA result,  the spin-degenerated $f_{0}$ state remains in the ground state even in the ferromagnetic phase: 
the $f_{0}$ state strongly hybridizes with the conduction state and does not show the spin-split. 
In order to obtain the appropriate ferromagnetic state, 
we must remove the degeneracy of $f_{0}$ spin states.
\section{LSDA+$U$ result}

\subsection{Choice of the $4f$ ground state}

According to the above discussion, 
we have chosen the fully spin- and orbital- polarized state $f_{0}^{up} \equiv \left| {l_z  = 0, s_x  = 1/2} \right\rangle$ as the $4f$ ground state in the following LSDA+$U$ calculation.
This choice may conflict  with the conventional idea, where the $4f$ ground state is described with the  strongly spin-orbit split $|j\rangle$ state with the comparatively smaller CEF splitting. 
In \crb, however, our choice is reinforced by the assumption that the $f_{0}$ 
state is stabilized by the larger CEF  splitting than the SOC splitting, due to the extremely short $c$ length.
The $\left| {j  = 5/2} \right\rangle$ state and $\left| {j  = 7/2} \right\rangle$ state 
hybridize well and compose the $|l\rangle$ and $|s\rangle$ states separately, in spite of the $|j\rangle$ states. 
Therefore we treated the  $|l_{z}\rangle$  and $|s_{x}\rangle$ states separately to construct the density matrix. 
The spin state is set as fully polarized along $x$ (or $y$) axis according to the direction of the observed magnetic moment. 
After constructing the initial density matrix where one electron occupies  $f_{0}^{up}$ state, 
the  effective potential shifts down the $f_{0}^{up}$ band and shift up the other $f$ bands.
Then the density matrix is calculated self-consistently so that  the final ground state is determined.

\subsection{Bandstructure and the magnetic moment}

\begin{figure}[htbp] 
\begin{center}
\includegraphics[width=7cm]{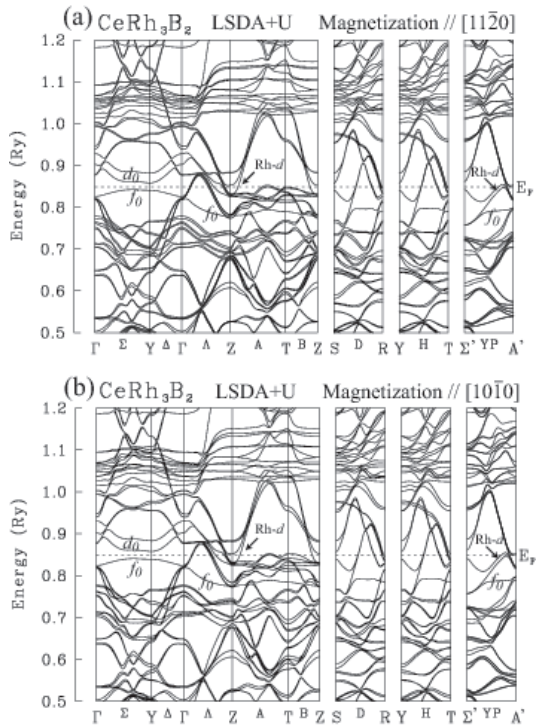}
\end{center}
\vspace{-4mm}
\caption{Calculated bandstructure of the ferromagnetic CeRh$_{3}$B$_{2}$  in the LSDA+$U$ scheme 
(a) with the [10\=10] magnetization  and (b) with the [11\=20] magnetization. 
The parameter $U$ is set as 0.3Ry. }
\label{fig:cerh.band.u}
\end{figure}

\begin{figure}[htbp] 
\begin{center}
\includegraphics[width=7cm]{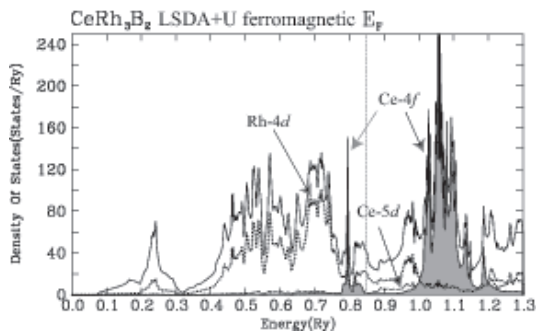}
\end{center}
\vspace{-4mm}
\caption{Calculated density of states for ferromagnetic CeRh$_{3}$B$_{2}$ in the LSDA+$U$ scheme 
with the [10\=10] magnetization. The notation is the same as in Fig.\ref{fig:cerh.dos.nonmag.lda}.}
\label{fig:cerh.dos.u}
\end{figure}

Figures \ref{fig:cerh.band.u} and \ref{fig:cerh.dos.u} show 
the calculated bandstructure and the density of states, respectively, in the LSDA+$U$ scheme where $U$ is set as 0.3 Ry.  
The Ce-$d$ and $f$ levels are artificially shifted upward by 0.13Ry and 0.12Ry respectively. 
This modification follows 
the previous FLAPW calculation for LaRh$_{3}$B$_{2}$\cite{larh.fermi}, 
where the La-$d$ and $f$ levels are shifted by 0.1Ry and 0.2Ry 
 to obtain the proper Fermi surfaces. 
Such modification is necessary to obtain the proper Fermi surfaces, as reported in 
LaB$_{6}$ \cite{eshift.1}, 
YbAl$_{3}$ \cite{eshift.2}, 
YNi$_{2}$B$_{2}$C \cite{eshift.3,eshift.4}, LuNi$_{2}$B$_{2}$C \cite{eshift.5}
 study, especially when a boron atom is involved in the system.

The LSDA+$U$ effective potential changes drastically the  bandstructure with respect to the LSDA result.
Near below the Fermi energy, the $f_{0}^{up}$ band is fully occupied as showing the large dispersion, which reflects the  symmetry of $Y_{30}$ wavefunction and the strong $f$-$f$ mixing. 
On the other hand, the spin-split $d_{0}^{up}$ and $d_{0}^{down}$ bands are fully unoccupied above the Fermi level.
Therefore only  Rh-$d$ bands cross the Fermi level, showing the large spin-splitting
at the Z point and along the YP axis. 
Comparing the bandstructure with the magnetization along $x$ (Fig. \ref{fig:cerh.band.u} (a)) and along $y$
(Fig. \ref{fig:cerh.band.u} (b)) axis, the  dependence of the bandstructure on the direction of magnetization is considerably small. 
The finally obtained $f^{1}$ ground state has not only the main component  of $\left| {l_z  = 0, s_x  = 1/2} \right\rangle$ state (as the initial set) but also the small components of $\left| {l_z  = \pm1,s_x  = 1/2} \right\rangle $ state 
because the SOC term  has the  non-zero matrix element between the different $\left| l_{z} \right\rangle$ states by $\pm 1$.
In consequence, the obtained ground state is described as 
\begin{equation}
\left\{ {a\left| {l_z  = 0} \right\rangle  + b\left| {l_z  =  \pm 1} \right\rangle } \right\}\left| {s_x  = 1/2} \right\rangle \nonumber 
\end{equation}
where $a=0.98$ and $b=-0.15$ are obtained. 
The mixing between different $\left|l_z\right\rangle$ state produces the orbital magnetic moment along the $x$ axis. 
As listed in Table \ref{tbl:cerh.moment}, with the comparison to the experimental result, 
the calculated  spin moment ($s_{x}=0.98\mu_{B}$; $s_{y}=0.90\mu_{B}$) of Ce-$f$  state is almost fully-polarized  and the orbital moment ($l_{x}=-0.85\mu_{B}$; $l_{y}=-0.81\mu_{B}$)  cancels the spin moment whereas Ce-$d$ and Rh-$d$ states don't have any significant magnetic moment.
Therefore, the net magnetic moment is only  0.11 (0.16)$\mu_{B}$/Ce along $x$ ($y$) axis.
This reduced moment is the key which explains 
the observed small saturated magnetic moment of 0.45 $\mu_{\it{B}}$.

\begin{table*}[t]
\caption{The orbital and spin magnetic moment ($\mu_{B}$) of ferromagnetic CeRh$_{3}$B$_{2}$:
The first three lines show the experimental result. The fourth and fifth lines show the calculated results with the magnetization is aligned along [11\=20] and [10\=10] axis, respectively.}
\label{tbl:cerh.moment}
\begin{scriptsize}
\begin{tabular}{lllllllllll} 
\hline
&\multicolumn{2}{c}{Ce 4f}&\multicolumn{2}{c}{Ce 5d}&\multicolumn{2}{c}{Rh-4d} &\multicolumn{2}{c}{total}\\ \cline{2-3} \cline{4-5} \cline{6-7} 
&\multicolumn{1}{c}{orbital}&\multicolumn{1}{c}{spin}&
\multicolumn{1}{c}{orbital}&\multicolumn{1}{c}{spin}&
\multicolumn{1}{c}{orbital}&\multicolumn{1}{c}{spin}&
\multicolumn{1}{c}{[11\=20]}&\multicolumn{1}{c}{[10\=10]}\\
\hline
Exp. $^a$ &  & &&&&& 0.447 & 0.451\\
Exp. $^b$ & 1.25 & -0.69&0.23&-0.41&&-0.03\\
Exp. $^c$ & 0.86 & -0.30&0.16&-0.34&&-0.05\\ \hline
Calc. {$s$//x}& -0.85  &  0.98 & 0.00&0.05&0.00&0.00&0.11\\
Calc. {$s$//y}& -0.81  &  0.90 & 0.01&0.06&0.00&0.00&&0.16\\
\hline
\\
\multicolumn{9}{l}{$^a$Magnetization measurement results from Ref.\cite{cerh.galatanu}.}\\
\multicolumn{9}{l}{$^b$Magnetic Compton scattering 
and  neutron diffraction results from Ref.\cite{cerh.mom1}.}\\
\multicolumn{9}{l}{$^c$Magnetic Compton scattering 
and neutron diffraction results from Ref.\cite{cerh.mom2}.}\\
\end{tabular}
\end{scriptsize}
\end{table*}


\subsection{Spin-split Fermi surfaces}

\begin{figure}[htbp] 
\begin{center}
\includegraphics[width=7cm]{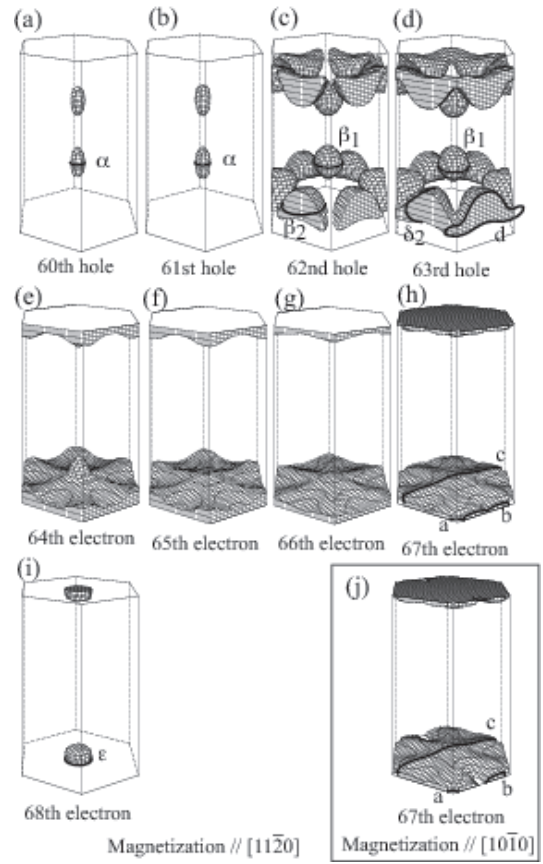}
\end{center}
\vspace{0mm}
\caption{Fermi surfaces of the ferromagnetic CeRh$_{3}$B$_{2}$ :
(a) hole surface from the 60th band, 
(b) hole surface from the 61st band,
(c) hole surface from the 62nd band,
(d) hole surface from the 63rd band,
(e) electron surface from the 64th band,
(f) electron surface from the 65th band,
(g) electron surface from the 66th band,
(h) electron surface from the 67th band and
(i) electron surface from the 68th band with the [11\=20] magnetization:
(j) electron surface from the 67th band band with the [10\=10] magnetization.}
\label{fig:cerh.lx.fermi}
\end{figure}
\begin{figure}[htbp] 
\begin{center}
\includegraphics[width=7cm]{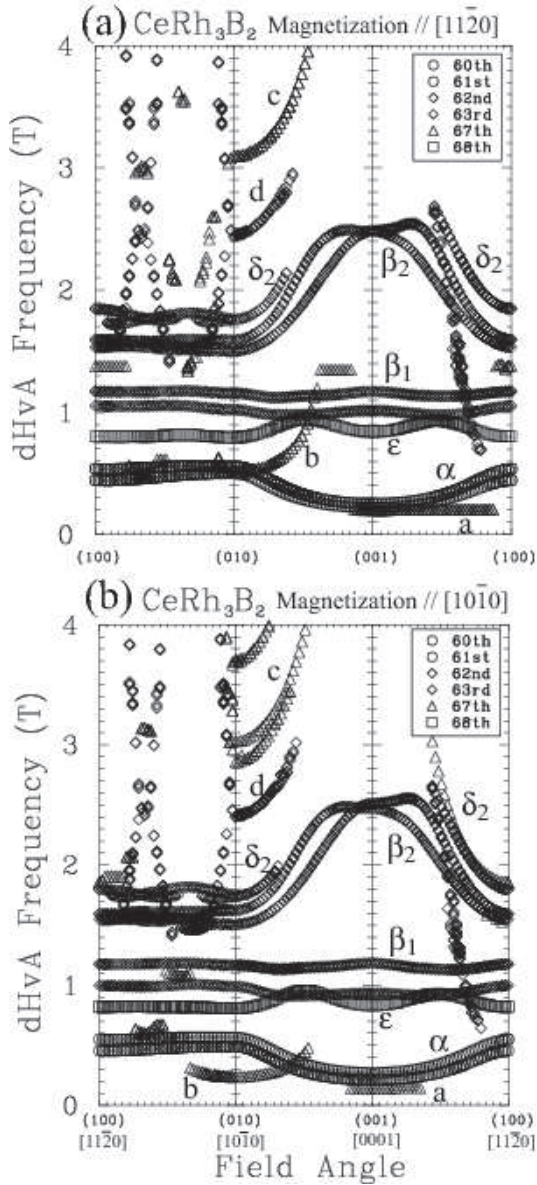}
\end{center}
\vspace{-4mm}
\caption{Angular dependence of the calculated dHvA frequencies 
of the ferromagnetic CeRh$_{3}$B$_{2}$ 
with (a) the [11\=20] magnetization and (b) the [10\=10] magnetization. The direction of the field is denoted with the orthorhombic lattice vector, together with the corresponding direction in the hexagonal lattice at the bottom of the figure.}
\label{fig:cerh.xydhva}
\end{figure}
In contrast to the nonmagnetic LaRh$_{3}$B$_{2}$, which has the five Fermi surfaces\cite{larh.fermi}, 
we have obtained the spin-split nine Fermi surfaces in ferromagnetic \crb \  as shown in Fig.  \ref{fig:cerh.lx.fermi}. 
The spin-splitting of the Fermi surfaces, which mainly consists of Rh-$d$ state, 
 is due to the magnetic contribution from the spin- and orbital-polarized $f$ ground state.
 We have also checked here the dependence of the Fermi surfaces on the magnetization axis. 
As changing the magnetization axis, it is found that only the 67th electron surface (shown in Fig. \ref{fig:cerh.lx.fermi} (h) and (j)) give the slightly  different shape.
In fact, when the magnetic field is applied to the ferromagnetic system and the magnetic moment is flipped,
 SOC changes the shape of Fermi surfaces according to the direction of the magnetic moment.
  
The angular dependence of the dHvA frequency derived from the Fermi surfaces 
is shown in Fig. \ref{fig:cerh.xydhva}. 
Although the shape of the Fermi surfaces can change progressively according to the applied magnetic field,  
we used rigid Fermi surfaces independent on it. 
Recalling that the magnetic moment is oriented in the basal plane, it may be sufficient to consider the Fermi surfaces only with the  [11\=20]  and [10\=10] magnetization.  
Between the two cases, the significant difference is  
due to the 67th electron surface whereas the rest show the similar frequencies and angle dependence.  
The dHvA branches $\alpha$, $\beta_{1}$, $\beta_{2}$, $\delta_{2}$ and $\varepsilon$ are 
named by the experimental results whereas the branches $a$, $b$, $c$ and $d$ are named here.
The dHvA frequencies and the electron mass are compared with the experimental results\cite{cerh.dhva}, as listed in Table \ref{tbl:cerh.mass}.
In the rest of this subsections, we closely look the comparison and discuss the spin-split effect on the Fermi surfaces. 
\begin{scriptsize}
\begin{table}
\caption{Experimentally observed dHvA frequencies $F_{\rm{exp}}$ and 
calculated dHvA frequencies $F_{\rm{calc}}$ in units of $10^{3}$T, 
cyclotron masses 
of $m_{\rm{exp}}$ {\&} $m_{\rm{calc}}$ in unit of the free electron mass 
and mass enhancement factor  $\lambda$. 
The magnetic field is applied along the [0001] and {[10\=10]} direction. 
The experimental value is taken from  Ref. \cite{cerh.dhva}.}
\label{tbl:cerh.mass}
\begin{scriptsize}
\begin{tabular}{lllllllll} \hline
\hline
Orbit & band & $F_{\rm{exp}}$& $F_{\rm{calc}}$ & $m_{\rm{exp}}$ & $m_{\rm{calc}}$ &$\lambda$          \\
&&($10^{3}$T)&($10^{3}$T) &($m_{0}$)&($m_{0}$)\\
\hline
[0001] \\
$\alpha$ & 60 & 0.15 & 0.20 & 0.33 & -0.12 & 1.8 \\
$\alpha$ & 61 & 0.15 & 0.26 & 0.37 & -0.14 & 1.6 \\
$\beta_{1} $ & 62  &0.84&0.92&0.60&-0.45&0.3\\
           		    & 63 &        &1.18&       &-0.63&    \\
$\beta_{2} $ & 62 & 2.14 & 2.48 & 2.0 & -0.91 & 1.2\\
$\gamma_{2}$ &    & 2.39 \\
$\varepsilon$   & 68 & 1.35 & 0.84 & 2.3 & 0.27 & 7.5 \\
{[10\=10]}\\
$\beta_{2} $ & 62 & 1.42 & 1.52 & 1.9& -0.54 & 2.5\\
                    &      &  1.59 & 1.62 &      & -0.50 & 2.8 \\
$\delta_{2} $ &63 & 2.12 & 1.72 & 1.4 & -0.55 &1.55\\
$\varepsilon $  & 68 & 0.97 & 0.80 & 1.7 & 0.35 & 3.86 \\
\hline
\\
\end{tabular}
\end{scriptsize}
\end{table}
\end{scriptsize}

The 60th and 61st ellipsoidal hole surfaces correspond to the 30th surface of LaRh$_{3}$B$_{2}$ 
where the splitting is relatively small. 
The two dHvA branches named $\alpha$ show the satisfactory agreement with the experimental result.\cite{cerh.dhva} 

The 62nd and 63rd hole surfaces correspond to the 31st surface of LaRh$_{3}$B$_{2}$. 
The closed spherical Fermi surfaces named $\beta_{1}$ shows similar aspect of the spin-splitting to the case of $\alpha$ surfaces, however, only one of the branches has been detected in the measurement. 
On the other hand, the large spin-splitting between the 62nd surface and the 63rd surface
 is essential to reproduce the experimental Fermi surfaces.   
The splitting separates the closed  62nd surfaces ($\beta_{2}$) and the connected 63rd surface ($\delta_{2}$) forming a ring along $k_{z}$ axis. 
It results in the fact that the $\beta_{2}$ dHvA branch is observed when the field is applied along [0001], whereas the  $\delta_{2}$ branch is not observed.
This result is at variance with the experimental expectation by  Okubo {\it et.al.}\cite{cerh.dhva}, 
where they have attributed 
the $\delta_{2}$ branch to the 64th-67th sheet-shape electron surfaces.

The 64th-67th electron surfaces have  typical quasi-one-dimensional flat shapes 
which correspond to the 32nd and 33rd surfaces of LaRh$_{3}$B$_{2}$.  
The 67th surfaces has the closed orbits, $a$, $b$ and $c$, 
however, the corresponding dHvA branches show the different angular dependence from the observed $\delta_{2}$ branch,
therefore we insist again on that $\delta_{2}$ is the spin-split counterpart of the $\beta_{2}$ branch.
 
The 68th ellipsoidal Fermi surface centered at the $Z$(0, 0, 1/2) point correspond to 
the 34th surface of LaRh$_{3}$B$_{2}$. 
The counterpart of the surface disappears due to the large spin splitting. 
The $\varepsilon$ branch show the similar angular dependence with the experimental result, 
however the value of the cross section  is  underestimated here.

To summarize the above discussion, we have obtained the proper Fermi surfaces, which explain well the experimentally observed dHvA branches.
The angular dependence of $\alpha$, $\beta_{1}$, $\beta_{2}$, $\delta_{2}$, $\varepsilon$ branches
 is substantially in good agreement with the experiment, 
while the  cross section  of $\delta_{2}$ and $\varepsilon$ branches are underestimated in the calculation (cfr. Table. \ref{tbl:cerh.mass}).

The missing branches in the calculation are  $\gamma_{1}$ and $\gamma_{2}$ branches,
observed with $H\parallel$ [11\=20] and with $H\parallel$ [0001], respectively.
In the experimental study,  Okubo {\it et.al.} has attributed both of them to the 64th-67th flat surfaces. 
As comparing the angular dependence, the calculated $b$ orbital may correspond to the observed $\gamma_{1}$ branch
whereas we cannot reproduce the $\gamma_{2}$ branch.
Although the angular dependence of the observed $\gamma_{2}$ branch implies that 
there is a large closed orbit perpendicular to $k_{z}$ axis,
it is hard to construct such a large orbit from these flat surfaces
without changing the volume of the electron surfaces, 
but the volumes for electrons and holes must be compensated. 
Therefore we guess the $\gamma_{2}$ branch originates from another surface, such as 63rd hole surface.
The hypothesis may be valid if one assumes larger spin-splitting between the 62nd and 63rd surfaces  so as to enhance the volume inside the 63rd hole surface. 
In such situation, the cross sectional area of hole orbit $\delta_{2}$ is increased and, at the same time, the area of  an hole orbit inside the ring-shaped surface is decreased.
The hole orbit with the small area causes the dHvA branch, which property may be consistent with the experimentally observed $\gamma_{2}$ branch. 
In this context, the cross sectional area of $d$ orbital is also enhanced and therefore the $d$ branch may be shifted upward,   out of the observable range. It explains why the calculated $d$ branch has not been observed. 

More serious problem is that the spin-splitting of $\beta_{1}$ branches have not been experimentally observed ($i. e.$ only one of the pair is observed). 
Since $\beta_{1}$ surfaces have the similar band character (mainly Ce-$d$ and B-$p$ character)
and the cross sectional area to those of $\alpha$ surfaces, 
the spin-splitting effect in dHvA frequency is supposed to be in proportion to the  cyclotron mass, 
which is observed approximately twice as large as that of $\alpha$ orbital.
Therefore it is supposed that $\beta_{1}$ branches must show the spin-splitting twice as large as $\alpha$ branches, as indeed shown in our calculation result. 
So far, there is no clear reason to explain why only the spin-splitting of $\alpha$ branch has been observed but not the one of $\beta_{1}$ branch. 

As we end this subsection, it should be emphasized again that 
the spin-splitting is important to discuss the 
observed dHvA result for  ferromagnetic CeRh$_{3}$B$_{2}$. 
The occupied $f_{0}^{up}$ band doesn't cross the Fermi level so that the 
 topology of the Fermi surfaces  is similar to that of LaRh$_{3}$B$_{2}$, 
however, the  hybridization from $f_{0}^{up}$ band, which is very close to Fermi level,  
causes the spin-splitting for the conduction bands.
In order to discuss the effect, let us have a careful look at the bandstructure. 
\begin{figure}[htbp]
\begin{center}
\includegraphics[width=7cm]{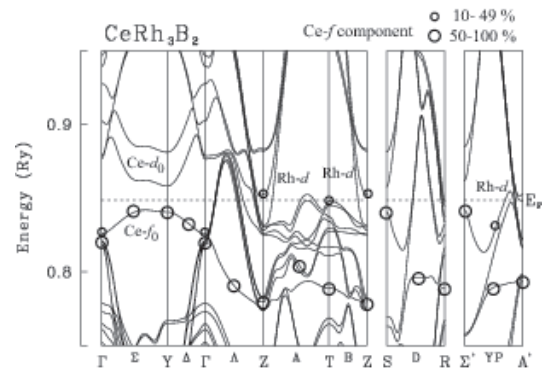}
\end{center}
\vspace{-4mm}
\caption{Bandstructure of the ferromagnetic CeRh$_{3}$B$_{2}$. 
The Ce-$f$ component is shown as round symbols  in the bands.  See the text for the details.}
\label{fig:cerh.enl.band}
\end{figure}
Figure \ref{fig:cerh.enl.band} shows the bandstructure of the ferromagnetic CeRh$_{3}$B$_{2}$ 
with the [11\=20] magnetization  within the enlarged energy scale.
The bands which have large Ce-$f$ component is indicated with round symbols, 
which size stands for the ratio of the component.
It is shown that the $cf$ interaction is large around the Z and T points and along YP axis, 
then the conductive Rh-$d$ band is largely spin-split.
The splitting in the vicinity of the Z point corresponds to the $\varepsilon$ surface splitting and 
the splitting along YP axis corresponds to the $\beta_{2}$-$\delta_{2}$ surface splitting.

\subsection{The effect of the choice of ground state}

In the above LSDA+$U$ calculation, we have chosen the $f_{0}^{up}$ state as the ground state, instead of the conventional CEF state. 
In order to check the validity of the consideration, 
another LSDA+$U$ calculation was  performed, 
where the ground state is chosen as a hexagonal CEF state  $\left| {j  = 5/2, j_z =1/2} \right\rangle$ state. 
In this choice, it is assumed that the Kramers pair  $\left| {j  = 5/2, j_z =\pm1/2} \right\rangle$ of the CEF level in hexagonal symmetry  is spin-split so that one of the pair is fully occupied.
The spin magnetization is set as parallel to $z$ axis ([0001] direction in the hexagonal symmetry). 

\begin{figure}[htbp] 
\begin{center}
\includegraphics[width=7cm]{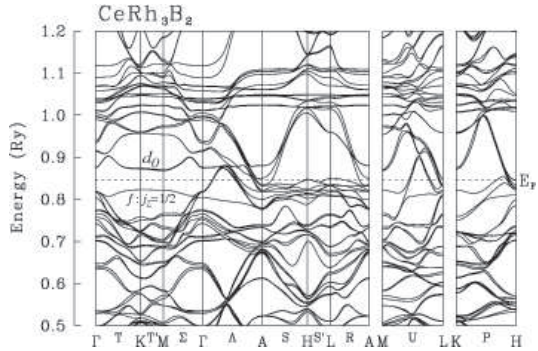}
\end{center}
\vspace{-4mm}
\caption{Calculated bandstructure of the ferromagnetic CeRh$_{3}$B$_{2}$  in the LSDA+$U$ scheme 
with the [0001] magnetization. 
The parameter $U$ is set as 0.3Ry. }
\label{fig:cerh.jz5.band}
\end{figure}
Figure \ref{fig:cerh.jz5.band} shows the 
calculated bandstructure of  
the ferromagnetic CeRh$_{3}$B$_{2}$ with $\left| {j  = 5/2, j_z =1/2} \right\rangle$ ground state with $U=0.3$ Ry.
In this calculation, the Ce-$d$ and $f$ levels are artificially shifted upward by 0.13 Ry and 0.08 Ry respectively as trying to fit the Fermi surfaces to the experimental dHvA result.
The calculated orbital and spin magnetic moment  
is 0.45 and -0.18 $\mu_{\rm B}$ along [0001] axis, 
which is compared with the expectation value of the CEF level as 
0.57 and -0.14 $\mu_{\rm B}$, respectively.
The topology of Fermi surfaces (not shown) reflects the hexagonal symmetry and
the dHvA frequency shows different angular dependence from the prior LSDA+$U$ result;   
the $\beta_{2}$ and $\delta_{2}$ branches are widely split, 
but the spin-splitting at $\alpha$ branch is not seen. 
This is fatal to reproduce the experimental Fermi surfaces so that the prior LSDA+$U$ calculation with $f_{0}^{up}$ ground state is favored.
Besides, the magnetic moment along [0001] axis of this ground state is totally different from the experimental result.

\section{Conclusion}

Here we briefly review the characteristic of Fermi surface at magnetic 4$f$ electron system as focusing on the localized vs itinerant behavior. 
It is known that spin-split Fermi surfaces are observed at magnetic Pr and Nd compounds, as explained by localized $f$ model. 
For example,  filled skutterudite NdFe$_{4}$P$_{12}$ shows  clear spin-pairs of dHvA branches which shows small spin-splitting, 
as compared to LaFe$_{4}$P$_{12}$.\cite{harima_ndfe4p12} 
This shows the localized character of 4$f$-electrons in Nd and the existence of small but sizable $c$-$f$ interaction. 
By contrast, Ce compounds, due to the rather stronger $c$-$f$ interaction, 
 are classified into several groups as following:
 
1) A localized model is valid at CeAl$_2$ and CeB$_6$,\cite{ceal2, ceal2b, ceb6} 
where the Fermi surfaces are quite similar to non-$f$ reference
system, LaAl$_2$ and LaB$_6$, as indicating that the $4f$ electron
is essentially localized and contribute little to the formation
of the Fermi surfaces.  

1') However, CeSb\cite{cesb.hillebrecht, cesb.kasuya} at ferromagnetic
phase shows the quite large spin-splitting of dHvA branches
(e.g. $\beta_1$, $\beta_2$, $\beta_3$ and $\beta_4$), where one cannot easily assign these
branches in pairs due to the various angular dependence, 
because anisotropic hybridization based on the local $f$-electron affects the Fermi surface topology.
 
2) At CeSn$_3$, the $4f$ electrons are considered as itinerant
electrons; The Fermi surfaces are well explained by band calculation,
where the $f$ electrons are treated in the same manner
as usual $s$, $p$, $d$ conduction electrons.\cite{cesn3.umehara}

Recently, heavy fermion superconductivity without spacial
inversion symmetry has been observed at CePt$_3$Si and
related compounds, Ce$TX_3$ ($T$=Rh, Ir, Co; $X$=Si, Ge).\cite{cept3si.bauer} 
It is theoretically expected that each Fermi surface of magnetic
CePt$_3$Si, $\alpha$, $\beta$ and $\gamma$ is spin-split by the Rashba-like antisymmetric
spin-orbit coupling.\cite{cept3si.samokhin}
Besides, even nonmagnetic reference
system, La$T$Ge$_3$ ($T$=Fe, Co, Rh, Ir) shows the split
Fermi surfaces.\cite{latge3.kawai} 
Then no further spin-split is observed in this system, even if it is magnetically ordered.
Such the large spin-orbit band splitting due
to the lack of the inversion symmetry is considered as the
origin of the particular symmetry of the superconducting gap node.

CeRu$_2$Si$_2$ shows metamagnetic transition from the nonmagnetic
ground state to the magnetic state at $H_m$$\sim$7.7T,
where both the occupied Ce-$f^1$ state and the magnetic moment
concomitantly change.\cite{ceru2si2.haen} 
A recent LSDA+$U$ calculation
of the Fermi surfaces with and without externally applied
magnetic field explains well the observed dHvA branches at
both states, showing that some of dHvA branches are largely
spin-split in the applied magnetic field.\cite{suzuki}
The transition is attributed to the change of Fermi surfaces from 2) to 1').

At Ce$T$In$_5$ ($T$=Co, Rh, Ir) series, known as heavy
fermion superconductors, FLAPW band calculations with
itinerant $4f$ model well explain the Fermi surfaces of CeIrIn$_5$
and CeCoIn$_5$,\cite{115.haga, 115.settai} whereas the Fermi surfaces of CeRhIn$_5$
shows the high deviation from them, which implies the localized
nature of the $4f$ electrons.\cite{115.shishido}
Therefore, the $f$electron behavior changes a lot even in the similar compounds of Ce$T$In$_5$ series.  

Here we note that CeRh$_{3}$B$_{2}$ has similar aspect to CeSb case.  
The above theoretical results compared with the experimental indicate that only LSDA+$U$ calculation  with $f_{0}^{up}$ ground state can explain the experimentally observed Fermi surfaces at ferromagnetic CeRh$_{3}$B$_{2}$. 
It is reasonable that due to the non-appropriate treatment of electron correlation, a bare LSDA calculation fails to reproduce the observed Fermi surfaces as well as the expected magnetic moment. 
We emphasize that at this system, the spin-splitting of the conduction band is strongly affected by the $c$-$f$ interaction so that 
it is not obvious to find the spin-split pairs of experimentally observed dHvA branches. 
Therefore LSDA+$U$ calculations are necessary to assign the dHvA branches to each Fermi surface,  
where we showed in this paper the ground state of Ce-$f^{1}$ state and the direction of the magnetic moment changes the shape of Fermi surfaces. 
Along this line, we can unambiguously specify the electronic state of 4$f$ electron system using by LSDA+$U$ method and comparing with the dHvA result. 
The validity of the $f_{0}^{up}$ ground state can be confirmed by other means of experiment: 
Indeed, a preliminary result of polarization-dependent photoelectron spectra recently shows the evident peak of $f^{0}$ state below the Fermi energy.\cite{Imada}

In summary of this paper, the Fermi surfaces of ferromagnetic CeRh$_{3}$B$_{2}$ is theoretically investigated.
We showed that the spin-splitting at the conduction band is caused by the magnetic contribution from fully spin-polarized Ce-$f_{0}$ band.  
By considering the $f_{0}$ ground state and the effect to the conduction bands, we explained well the observed small magnetic moment as well as the experimental dHvA result.
The large band dispersion of $f_{0}$ state due to  the $f$-$f$ direct mixing between Ce sites may give a hint on the origin of the High $T_\textrm{C}$ at CeRh$_{3}$B$_{2}$.

\acknowledgments

Authors thank Y. \=Onuki and  H. Katayama-Yoshida for the helpful discussions and comments.
This work is partly supported by 21st Century COE from the Ministry of Education, Culture, Sports, Science and Technology.

\bibliographystyle{junsrt}

\begin{thebibliography}{1}
\bibitem{heavy1}
Y. \=Onuki, T. Goto, T. Kasuya:  
Materials Science and Technology, Materials Science and Technology ed K.H.J. Buschow (VCH,  Weinheim ) ${\bf 3A}$ (1991) 545.

\bibitem{heavy2}
Y. \=Onuki, A. Hasegawa:  
Handbook on the Physics and Chemistry of Rare Earths (Elsevier, Amsterdam), ${\bf 20}$  (1995) Chap 135, 49. 

\bibitem{magphase1}
T. Tsuchida and Y. Nakamura: 
J. Phys. Soc. Japan ${\bf 22}$ (1967) 942. 

\bibitem{magphase2}
H. Bartholin, D. Florence, Wang Tcheng-si and O. Vogt: 
Phys. Stat. Sol. (a) ${\bf 24}$ (1974) 631.

\bibitem{multipole}
J.M. Effantin, J. Rossat-Mignod, P. Burlet, H. Bartholin, S. Kunii, T. Kasuya: 
J. Magn. Magn. Mater. ${\bf 47-48}$ (1985) 145.

\bibitem{cerh.curie1} 
S. K. Dhar, S. K. Malik and R. Vijayaraghavan: 
J. Phys. C: Solid State Phys.
${\bf 14}$ (1981) L321.

\bibitem{cerh.galatanu}
A. Galatanu, E. Yamamoto, T. Okubo, M. Yamada and A. Thamizhavel: 
J. Phys. : Condens. Matter
${\bf 15}$ (2003) S2187.

\bibitem{cerh.kasuya.okabe}
M. Kasaya, A. Okabe, T. Takahashi, T. Satoh and T. Kasuya: 
J. Magn. Magn. Mater. 
${\bf 76-77}$ (1988) 347.

\bibitem{cerh.f.it}
S. K. Malik, R. Vijayaraghavan, W. E. Wallace and S. K. Dhar: 
J. Magn. Magn. Mater. 
${\bf 37}$ (1983) 303. 

\bibitem{cerh.d.it}
S. K. Dhar, S. K. Malik and R. Vijayaraghavan:
J. Phys. C 
${\bf 14}$ (1981) L321.

\bibitem{cerh.f.loc1}
A. L. Cornelius, J. S. Schilling and R. N. Shelton: 
Phys. Rev. B 
${\bf 49}$ (1994) 3995.

\bibitem{cerh.f.loc2}
K. Yamaguchi, H. Namatame, A. Fujimori, T. Koide, T. Shidara, M. Nakamura, A. Misu, H. Fukutani, M. Yuri, M. Kasaya, H. Suzuki and T. Kasuya: 
Phys. Rev. B ${\bf 51}$ (1995) 13952. 

\bibitem{cerh.f.loc3}
E. V. Sampathkumaran, G. Kaindl, C. Laubschat, W. Krone and G. Wortmann: 
Phys. Rev. B ${\bf 31}$ (1985) 3185.

\bibitem{cerh.f.loc4}
S. A. Shaheen, J. S. Schilling and R. N. Shelton: 
Phys. Rev. B ${\bf 31}$ (1985) 656.

\bibitem{cerh.f.loc5}
A. Fujimori, T. Takahashi, A. Okabe, M. Kasaya and T. Kasuya: 
Phys. Rev. B ${\bf 41}$ (1990) 6783.

\bibitem{cerh.f.loc6}
J. A. Alonso, J. X. Boucherle, F. Givord, J. Schweizer, B. Gillon and P. Lejay: 
J. Magn. Magn. Mater. ${\bf 177-181}$ (1998) 1048.

\bibitem{cerh.dhva}
T. Okubo, M. Yamada, T. Thamizhavel, S. Kirita, Y. Inada, R. Settai, H. Harima, K. Takegahara, A. Galatanu, E. Yamamoto and Y. \=Onuki: 
J. Phys.: Condens. Matter ${\bf 15}$ (2003) L721.

\bibitem{larh.fermi}
H. Harima and K. Takegahara: 
J. Magn. Magn. Mater. ${\bf 272-276}$ (2004) 475.

\bibitem{flapw}
E. Wimmer, H. Krakauer, M. Weinert and A. J. Freeman:
Phys. Rev. B ${\bf 24}$ (1981) 864.

\bibitem{koellingharmon}
D. D. Koelling and B. N. Harmon: 
J. Phys. C: Solid State Phys. ${\bf 10}$ (1977)  3107. 

\bibitem{kubler}
J. K\"ubler: 
{\it Theory of Itinerant Electron Magnetism} (Oxford Science Publications, New York, 2000).

\bibitem{anisimov}
A. I. Liechtenstein, J. Zaanen and V. I. Anisimov:
Physical Review B ${\bf 52}$ (1995) R5467.

\bibitem{harima.ldau}
H. Harima:
J. Mag. Magn. Matter ${\bf 83-84}$ (2001) 226.

\bibitem{suzuki}
M. -T. Suzuki, and H. Harima: 
to be published in J. Phys. Soc. Jpn.

\bibitem{cerh.crysl}
H. C. Ku, G. P. Meisner, F. Acker and D. C. Johnston: 
Solid State Commun. 
${\bf 35}$ (1980) 91. 


\bibitem{cerh.apw}
K. Takegahara, H. Harima and T. Kasuya: 
J. Phys. Soc. Jpn. 
${\bf 54}$ (1985) 4743.

\bibitem{eshift.1}
H. Harima, O. Sakai, T. Kasuya and A. Yanase: 
Solid State Commun. 
${\bf 66}$ (1988) 603.

\bibitem{eshift.2}
T. Ebihara, Y. Inada, M. Murakawa, S. Uji, C. Terakura, T. Terashima, E. Yamamoto, Y. Haga, Y. \={O}nuki and H. Harima: 
J. Phys. Soc. Jpn. 
${\bf 69}$ (2000) 895.

\bibitem{eshift.3}
K. Yamauchi, H. Katayama-Yoshida, A. Yanase and H. Harima: 
Physica C 
${\bf 412-414}$ (2004) 225.

\bibitem{eshift.4}
K. Yamauchi and H. Harima: 
Physica B
${\bf 359}$ (2005) 597.

\bibitem{eshift.5}
Y. Nagai, Y. Kato, N. Hayashi, K. Yamauchi and H. Harima:
Phs. Rev. B
${\bf 76}$ (2007) 214514. 

\bibitem{cerh.mom1}
Y. Sakurai, M. Ito, J. Tamura, S. Nanao, A. Thamizhavel, Y. Inada, A. Galatanu, E. Yamamoto and Y.\={O}nuki: 
J. Phys.: Condens. Matter		
${\bf 15}$ (2003) S2183.

\bibitem{cerh.mom2}
A. Yaouanc, P. Dalmas de R\'{e}otier, J-P. Sanchez, Th. Tschencher and P. Lejay: 
Phys. Rev. B
${\bf 57}$ (1998) R681.


\bibitem{harima_ndfe4p12}
H. Sugawara, Y. Abe, Y. Aoki, H. Sato, 
M. Hedo, R. Settai, Y. \={O}nuki and H. Harima:
J. Phys. Soc. Jpn. 
${\bf 69}$ (2000) 2938. 

\bibitem{ceal2}
M. Springford and P. H. P. Reinders: 
J. Magn. Magn. Mater. ${\bf 76-77}$, (1988) 11.

\bibitem{ceal2b}
G. G. Lonzarich:
 J. Magn. Magn. Mater. $76-77$, (1988) 1.

\bibitem{ceb6}
Y. \={O}nuki, T. Komatsubara, P. H. P. Reinders, and M. Springford: 
J. Phys. Soc. Jpn. ${\bf 58}$, (1989) 3698. 

\bibitem{cesb.hillebrecht} 
F. U. Hillebrecht, W. Gudat, N. Martensson, D. D. Sarma, and M. Campagna: 
J. Magn. Magn. Mater. ${\bf 47-48}$, (1985) 221.

\bibitem{cesb.kasuya}
T. Kasuya, O. Sakai, J. Tanaka, H. Kitazawa, and T. Suzuki:
J. Magn. Magn. Mater. ${\bf 63-64}$, (1987) 9.

\bibitem{cesn3.umehara}
I. Umehara, Y. Kurosawa, N. Nagai, M. Kikuchi, K. Satoh, and Y. \={O}nuki: 
J. Phys. Soc. Jpn. ${\bf 59}$, (1990) 2848. 

\bibitem{cept3si.bauer}
E. Bauer, G. Hilscher, H. Michor, Ch Paul, E. W. Scheidt, A. Gribanov, Yu Seropegin, H. No\"el H, M. Sigrist and 
P. Rogl: 
Phys. Rev. Lett. ${\bf 92}$ (2004) 027003. 

\bibitem{cept3si.samokhin}
K. V. Samokhin, E. S. Zijlstra, and S. K. Bose:
Phys. Rev. B. ${\bf 69}$, (2004) 094514. 

\bibitem{latge3.kawai}
T. Kawai, H. Muranaka, M.-A. Measson, T. Shimoda, Y. Doi, T. Matsuda, Y. Haga, G. Knebel, G. Lapertot, D. Aoki, J. Flouquet, T. Takeuchi, R. Settai, and Y. \={O}nuki: 
J. Phys. Soc. Jpn. ${\bf 77}$ (2008) 064717.

\bibitem{ceru2si2.haen}
P. Haen, J. Flouquet, F. Lapierre, P. Lejay, and G. Remenyi: 
J. Low Temp. Phys. ${\bf 67}$ (1987) 391. 

\bibitem{115.haga}
Y. Haga, Y. Inada, H. Harima, K. Oikawa, M. Murakawa, H. Nakawaki, Y. Tokiwa, D. Aoki, H. Shishido, S. Ikeda, N. Watanabe and Y. \={O}nuki: 
Phys. Rev. B ${\bf 63}$ (2001) 060503.

\bibitem{115.settai}
R. Settai, H. Shishido, S. Ikeda, Y. Murakawa, M. Nakashima, D. Aoki, Y. Haga, H. Harima and Y. \={O}nuki: 
J. Phys.:	Condens. Matter ${\bf 13}$ (2001) L627.

\bibitem{115.shishido}
H. Shishido, R. Settai, D. Aoki, S. Ikeda, S. Araki, M. Nakashima, Y. Inada, Y. Haga, H. Harima, Y. Aoki, T. Namiki, H. Sato and Y. \={O}nuki: 
J. Phys. Soc. Jpn. ${\bf 71}$ (2002) Suppl. 276.

\bibitem{Imada}
S. Imada: private communication. 



\end{thebibliography}

\end{document}